\documentclass[aps,amsmath,amssymb,reprint, prl, twocolumn, superscriptaddress, nobibnotes]{revtex4}

\usepackage[utf8]{inputenc}
\usepackage{amsfonts}
\usepackage{bbold}
\usepackage{mathrsfs}
\usepackage{color}
\usepackage{epstopdf}

\usepackage{graphicx}
\usepackage{bm}
\usepackage{natbib}
\usepackage{graphicx}
\usepackage[textsize=tiny]{todonotes}
\usepackage[colorlinks=true, allcolors=blue]{hyperref}

\newcommand{\hh }[1]{ \hat{\bm{#1}} }
\newcommand{\m }[1]{ \mathbf{#1} }

\def\eq#1{{Eq.(\ref{#1})}}   
\def\fig#1{{Fig.\ref{#1}}}

\begin{document}

\title{
Surfing and crawling macroscopic active particles under \textcolor{black}{strong} confinement -- inertial dynamics}








 
\author{Marco Leoni}
\thanks{equally contributing authors}
\affiliation{Universit\'e Paris-Saclay, CNRS, IJCLab, 91405, Orsay, France}

\author{Matteo Paoluzzi}
\thanks{equally contributing authors}
\affiliation{Departament de Física de la Mat\`eria Condensada, Universitat de Barcelona, C. Martí Franqu\`es 1, 08028 Barcelona, Spain}
\affiliation{ISC-CNR,  Institute  for  Complex  Systems,  Piazzale  A.  Moro  2,  I-00185  Rome, Italy}  \affiliation{
Dipartimento di Fisica, Sapienza Universit\`a di Roma, Piazzale A. Moro 2, I-00185, Rome, Italy
}

\author{Sarah Eldeen}
\affiliation{Department of Physics, California State University Fullerton, CA 92831 USA}

\author{Anthony Estrada}
\affiliation{Department of Physics, California State University Fullerton, CA 92831 USA}

\author{Lauren Nguyen}
\affiliation{Department Chemistry and Biochemistry, California State University Fullerton, CA 92831 USA}

\author{Maria Alexandrescu}
\affiliation{Troy High School, Fullerton, CA 92831 USA}

\author{Karin Sherb}
\affiliation{Troy High School, Fullerton, CA 92831 USA}

\author{Wylie W. Ahmed}
\email[correspondence: ]{wahmed@fullerton.edu }
\affiliation{Department of Physics, California State University Fullerton, CA 92831 USA}

\date{\today}
\begin{abstract}

We study two types of active (self-propelled) macroscopic particles under confinement:~camphor surfers and hexbug crawlers, using a combined experimental, theoretical, and numerical approach.  Unlike widely studied microscopic active particles and swimmers, where thermal forces are often important and inertia is negligible, our macroscopic particles exhibit complex dynamics due expressly to active non-thermal noise combined with inertial effects.  \textcolor{black}{Strong} confinement induces accumulation at a finite distance \emph{within} the boundary and gives rise to three distinguishable dynamical states; both depending on activity and inertia.  These surprisingly complex dynamics arise already at the single particle level --- highlighting the importance of inertia in macroscopic active matter.


\end{abstract}

\maketitle

\section{Introduction}
Active matter is a rapidly growing field of research that studies the behavior of self-driven entities that exhibit rich dynamics and collective phenomena in systems covering a wide range of length scales~\cite{marchetti2013hydrodynamics,ramaswamy2010mechanics,bechinger2016active}.  This ranges from molecular-scale systems such as driven biopolymers~\cite{sanchez2012spontaneous} up to meter-scale systems like dense crowds of people~\cite{bain2019dynamic,bottinelli2016emergent,silverberg2013collective}.  Physical confinement
of active systems, critical not only to understand the effect of boundaries but also for applications to real-life systems,
triggers interesting dynamical behaviors, e.g.~collective motion, accumulation, segregation, phase separation, and freezing/fluidization~\cite{designe2010,vladescu2014filling,das2018confined,caprini2019activity,tailleur2009sedimentation, fily2014dynamics, yang2014aggregation, takatori2016acoustic}. 
While experimental observations in 
starling flocks have shown the importance of inertial effects in flocking transitions~\cite{attanasi2014information}, how inertia modifies some peculiar features of active systems, such as accumulation at the boundaries or anomalous density fluctuations remains unknown. 

For a system in thermal equilibrium the
density
follows
 the Boltzmann distribution \cite{huang2009introduction}. This means, for instance, that a Brownian particle 
in a container
 will uniformly explore the accessible space, without developing  currents. In stark contrast, systems of active particles exhibit a steady-state density profile with an accumulation peak at the confining wall~\cite{tailleur2009sedimentation,wagner2017steady,maggi2015multidimensional,marconi2016velocity,PhysRevResearch.2.023207,dauchot2019dynamics,marini2017pressure} as it has been observed in experiments~\cite{vladescu2014filling, takatori2016acoustic}.  
 Most studies have focused on overdamped systems due to the ubiquity of experimental work at the microscopic scale~\cite{bechinger2016active}, however, a  growing number of experimental observations highlight the importance of inertia in macroscopic active matter systems~\cite{dauchot2019dynamics,scholz2018inertial,deblais2018boundaries, bourgoin2019kolmogorovian,scholz2018rotating}.

\begin{figure}[t!]
\centering
\includegraphics[width=0.46\textwidth]{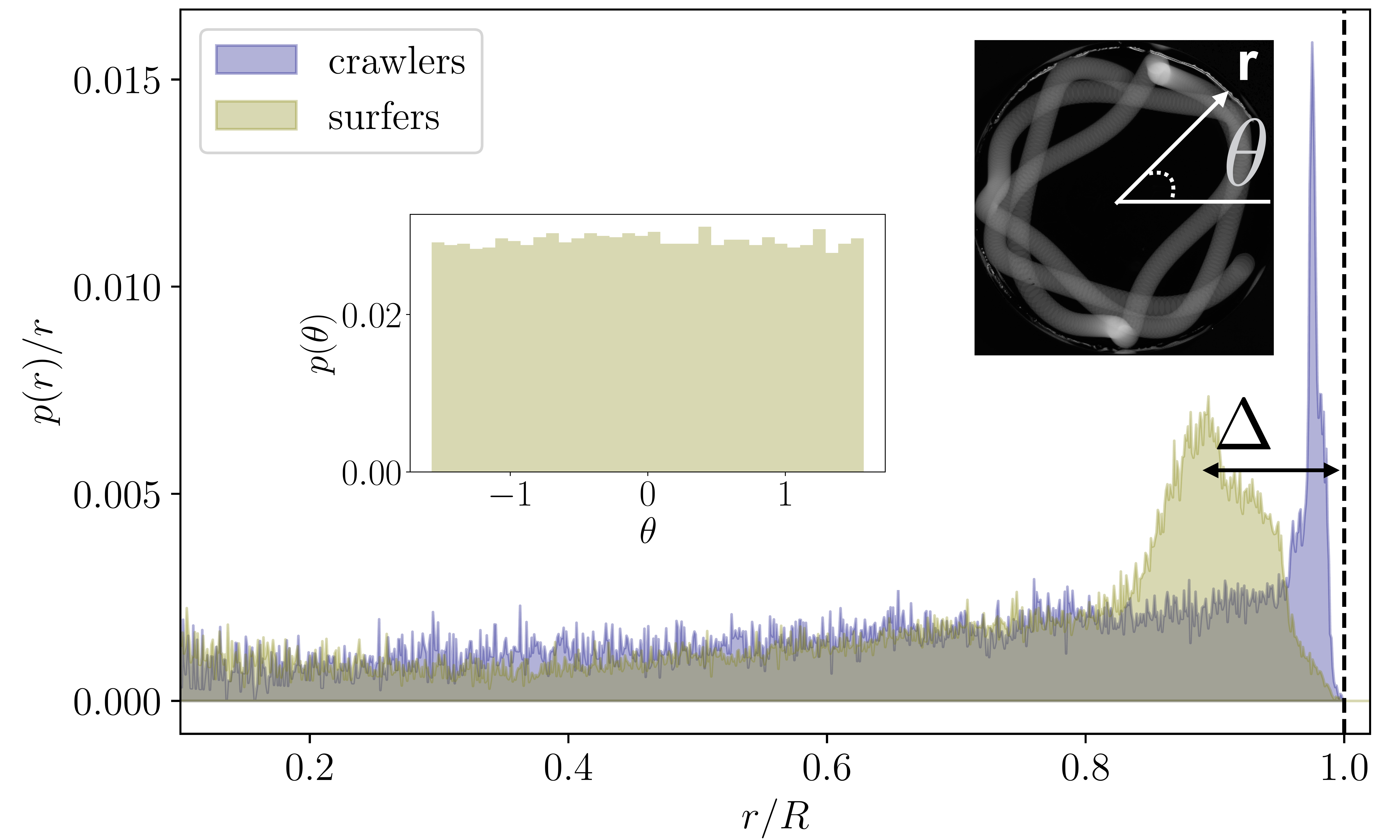}
\caption{Probability distribution of the position of an isolated active particle in presence of \textcolor{black}{strong} boundaries for camphor surfers (green) and hexbug crawlers (violet).
These distribution are peaked at a distance $\Delta$ within the system boundary (dashed vertical line), unlike in overdamped systems. 
Insets:
a representative angular probability distribution (left, units in radians) and a the time-lapse of a surfer's trajectory (right).
}
\label{fig:radial-distribution}
\end{figure}

In this work we study the role of a
strong confining boundary on the dynamics of single macroscopic active particles, at two length scales (mm and cm) where
inertia is non-negligible. We use two different systems, comprising both wet and dry active matter: camphor surfers, which glide at the fluid air interface via a surface tension-driven motion~\cite{boniface2019self}; and Hexbug crawlers, which are propelled 
on a solid surface  by a vibrating motor~\cite{dauchot2019dynamics}; hereafter, referred to as surfers and crawlers. In both systems, a single surfer or crawler is confined to a circular container with rigid walls.
Strikingly, due to the
interplay of
inertial dynamics and \textcolor{black}{strong} confinement we observe rich dynamics already at the single particle level. We observe:~(1) steady-state density distributions that exhibit an accumulation peak at a finite distance \emph{within} the confining wall (\fig{fig:radial-distribution}), and (2) transitions between three dynamical states that we call ``orbits'', ``epicycles'', and ``collisions'' (\fig{fig:fig4-transitions}).  Through experiments, modeling, and simulations we show that the self propulsion speed of the inertial active particles controls the location of the density peak and drives transitions between the three dynamical states.  Including inertia in models is critical in capturing the observed dynamics.


\section{Materials and Methods}

\subsection{Experimental systems}
Experiments consist of two separate systems: (1) Millimeter-scale camphor surfers were created by infusing agarose gel disks with camphor solution as studied previously~\cite{boniface2019self, soh2008dynamic}.  The resulting self-propelled surfer has a radius of $\sim3$ mm and a mass of $\sim40$ mg.  The dynamics is then studied by placing the surfer at the water-air interface in a circular petri dish of 10 cm diameter with 20 g of ultrapure water.  Self-propulsion is driven by gradients in surface tension.  (2) Centimeter-scale crawlers were created using a Hexbug nano~\cite{dauchot2019dynamics} that is trapped under an \textcolor{black}{inverted rigid isotropic cup} of radius $\sim3$ cm and a combined mass of $\sim10$ g.  \textcolor{black}{The Hexbug is 4.5 cm in length, which allows it to move freely within the cup and collide with its walls --- this combined Hexbug and cup system is referred to as the crawler.}  The crawler is then placed on a flat circular table of $\sim 1$ m diameter that has a vertical wall around its edge.  Self-propulsion is driven by a mechanical vibrating motor. Thus these two systems comprise wet and dry active matter systems and differ in scale by an order of magnitude. For both systems motion was recorded at 20-30 Hz.  In both systems, the active particle is free to move in-plane (\fig{fig:Exp_SM}) but experiences a vertical wall at the boundary.  For a surfer, collision with the boundary is likely mediated through capillary effects~\cite{soh2008dynamic}.  For a crawler, the collision with the boundary is mediated directly through physical contact of the cup with the wall.  \textcolor{black}{Therefore, both surfers and crawlers experience \textcolor{black}{strong} confinement to a circular container of size $R = 5$ cm and 50 cm, respectively, and exhibit a typical speed of $v_{exp} \sim 9$ cm/s.  This defines a typical timescale to traverse the container, $\tau = R/v_{exp}$, as $\sim 0.5$ s for surfers and $\sim 5.5$ s for crawlers.}

\subsection{Image capture and analysis}

\begin{figure}[t!]
    \centering
    \includegraphics[width=0.5\textwidth]{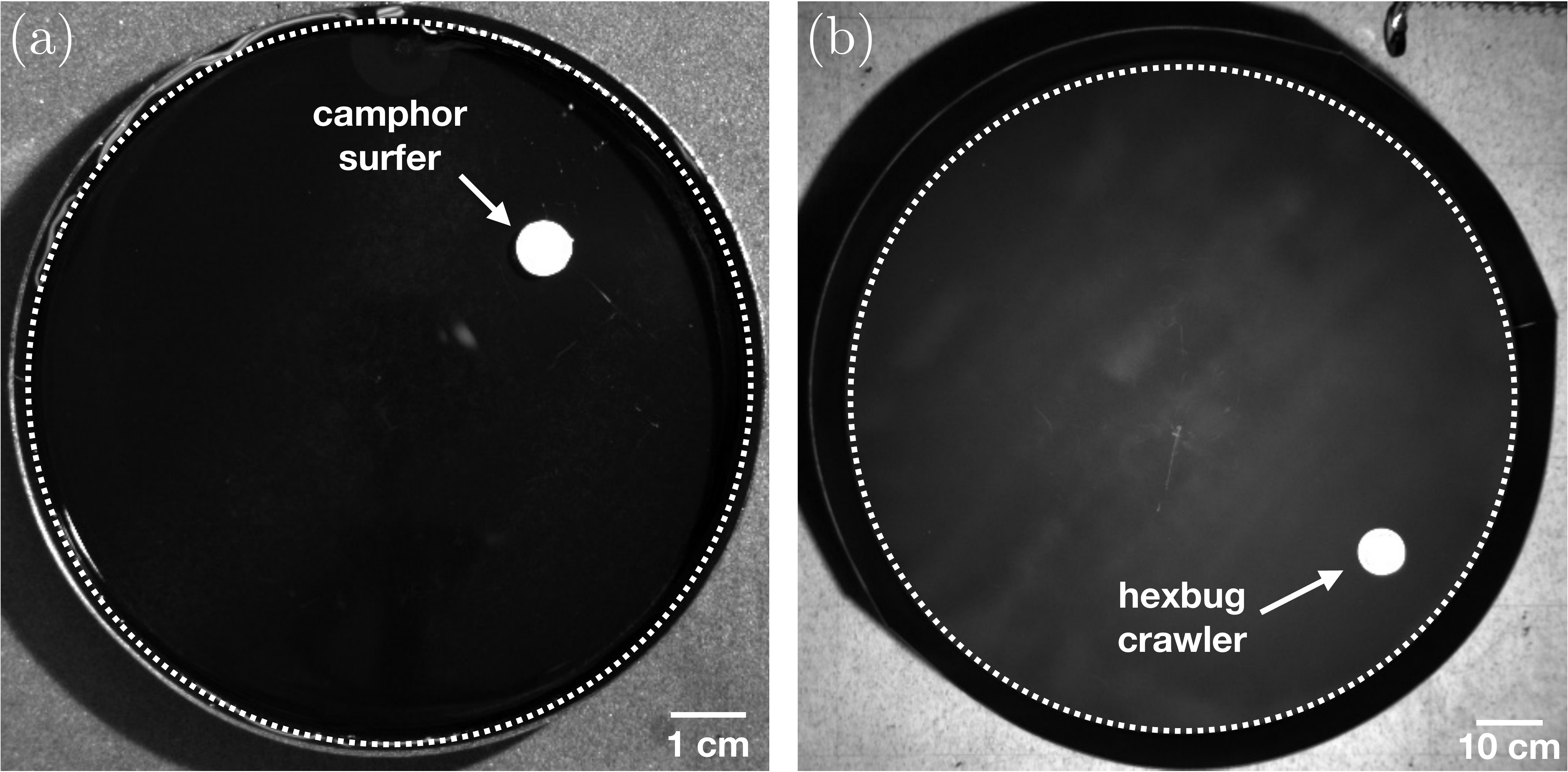}
    \caption{Representative images of camphor surfer (a) and hexbug crawler (b).  \textcolor{black}{Dashed line indicates boundary of container}.}
    \label{fig:Exp_SM}
\end{figure}

Representative images of a camphor surfer and hexbug crawler are shown in \fig{fig:Exp_SM}, where the dotted line indicates the container boundary. In both experiments the container was painted with anti-reflective black paint to enhance contrast and images were captured using identical CMOS cameras (Basler acA3088-57um, from Graftek Imaging) where 4x pixel binning was used at the time of acquisition resulting in an image of 768x516 pixels and saved as individual  linearly-encoded TIFF files.  A different lens for each type of experiment was used to accommodate the difference in scale: (1) Computar M3Z1228C-MP for surfers and (2) Kowa LMVZ4411 for crawlers --- both purchased from Graftek Imaging.  Captured image sequences were analyzed in MATLAB to determine particle trajectories using a custom-written image processing code. Briefly, images were thresholded and background noise was removed via filtering, and the centroid of the single particle was recorded for each frame.  Working with single macroscopic particles that remain in-plane and exhibit large contrast with their background is relatively straight-forward and thus the centroid of the particle could be consistently determined in every single frame.  Tracking precision was determined to be $\sim$1 pixel for both surfers and crawlers, resulting in an uncertainty of 0.2 mm and 0.2 cm, respectively.  This tracking precision corresponds to $\sim1/30$th of the particle diameter.

\subsection{Numerical Model} \label{num}
For modelling the dynamics of both, surfers and crawlers, 
we consider the following 
stochastic dynamics for the translational and rotational degrees of freedom  
\begin{eqnarray} \label{num_model}
m \ddot{\mathbf{r}} &=& - \gamma \dot{\mathbf{r}} + \gamma v_0 \mathbf{n} + \mathbf{f} \\
I \ddot{\varphi} &=& - \gamma_R \dot{\varphi} + \gamma_R \sqrt{2 D_r} \eta \; .
\end{eqnarray}
In Eq.~(\ref{num_model}), $\mathbf{r}=(x(t),y(t))$ is the position of  
the center of mass of the macroscopic disc at time $t$.
$m$ is the mass of the disc, $I$ the moment of inertia,
$\gamma$ and $\gamma_R$ the translational and rotational friction coefficients. 
The moment of inertia is due to the fact that in the underdamped regime we have to take into account
also the finite size of the particle. 
The particle is self-propelled along the direction given by the vector $\mathbf{n}=(\cos \varphi(t), \sin \varphi(t))$. $v_0$ is the self-propulsion speed, $\eta$ is a random noise that satisfies 
$\langle \eta \rangle =0 $ and $\langle \eta(t) \eta(s) \rangle = \delta(t-s)$, and $D_r$ is the rotational diffusion coefficient. 
In writing Eq.~(\ref{num_model}), we consider as negligible the random fluctuations acting on translational degrees of freedom. 
This is motivated by the fact that, in our experiments, the self-propulsion is the leading stochastic force acting on the system, i.~e., the diffusion due to the thermal bath is orders of magnitude smaller than the displacement due to the self-propulsion.
Since we are considering the system confined by a circular container,
$\mathbf{f}$ is the force exerted by the boundaries on the particle. 
In particular, the boundaries are modelled using 
a conservative central 
potential $\phi(r_{wp})=(A/r_{wp})^{12}$ with $A$ a coupling constant that fixes the equilibrium distance between the particle and the wall. We consider a
circular container of radius $R$ \cite{PhysRevLett.115.188303,paoluzzi2020narrow}. $r_{wp}$ indicates
the wall/particle distance.
%

\begin{figure}[t]
\centering
\includegraphics[width=0.47\textwidth]{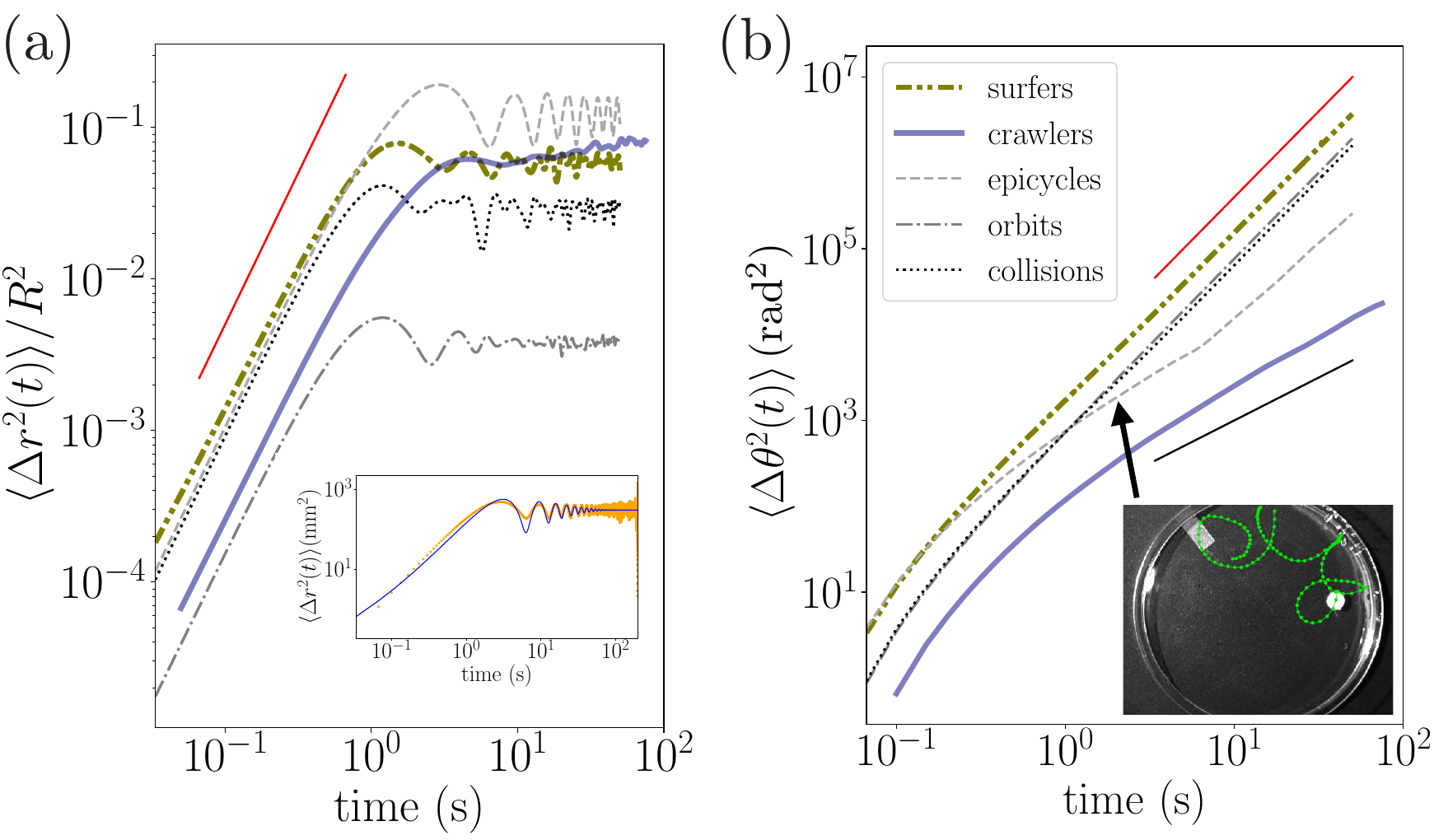}
\caption{
Mean square displacements (MSDs).
(a) MSD for the radial coordinate, $r(t)$, normalized to the container size, $R$. A crossover from ballistic to a flat plateau indicates confinement. In most experimental runs oscillatory behaviour is observed (inset, shows theoretical fit for surfers). From this fit we estimate $m/\gamma \sim  40$ and 5 s for surfers and crawlers, respectively.
(b) MSD for the angular coordinate 
$\theta(t)$.  
The angular
dynamics is consistent with ballistic motion in surfers and diffusive  motion for crawlers. Kinks, describing multiple timescales, are observed e.g.~during epicycles (inset).  \textcolor{black}{Data shown for epicycles, collisions, and orbits are from surfers.} Power laws of ballistic (solid red) and diffusive (solid black) dynamics are shown as a guide for the eye. 
} 
\label{fig:MSDs}
\end{figure}

\section{Results and Discussion}
\subsection{Spatial distributions and MSD}
First, we focus on the average properties
of  active particles interacting with the boundaries in presence of both inertia and friction. 
To characterize the statistical behaviour of the system, we tracked the positions (using custom code written in MATLAB), computed the polar coordinates $\mathbf{r}(r,\theta)$ with origin at the center of the container, and calculated the probability distributions of their positions, $p(r)$ and $p(\theta)$~(\fig{fig:radial-distribution}). 
As expected $p(\theta)$ shows a uniform distribution because the containers are rotationally symmetric about their origin. 
The uniform distribution in angle likely arises from the rotating dynamics 
of particles along the boundary of the container, see below for details,
but is also consistent with random motion.
Interestingly, the radial distribution, $p(r)$,
 of these active particles under confinement is new: 
 it is not uniform, as one might expect in the case of Brownian particles,
nor peaked at the boundary, 
as it was found  
for active Brownian particles  --- e.g. micro-swimming bacteria~\cite{vladescu2014filling}.
Instead, here, the most probable configuration 
is at some finite distance \emph{within} the boundary of the container as shown in \fig{fig:radial-distribution} by $\Delta$.  This observation, along with our analytic and numerical models showing $\Delta$ depends on activity, $v_0$, is the first main result of this work.

To gain insight on this behaviour,
we computed the radial and
angular mean squared displacement (MSD), as shown in \fig{fig:MSDs}.
For both surfers and crawlers the radial MSD initially
grows ballistically in time followed by a crossover to a flat plateau as expected for the motion of a particle confined to a circular region.  This suggests the persistence length of a surfer/crawler is greater than our system size.
Often oscillations are observed for surfers 
in the crossover region \fig{fig:MSDs}(left). 
The angular displacement instead is simpler:  typically   ballistic  for surfers and nearly diffusive for crawlers. 
Sometimes kinks and  crossovers between two 
regions with similar slope are observed. We interpret these as signatures of a multi-scale dynamics: for instance local rotational motion --- \fig{fig:MSDs}(right), see inset  ---
followed by an overall large scale rotation along the boundary. 

An important difference with respect to microscopic systems
is that these particles are too large to be sensitive to thermal fluctuations. However, we still observe noise, e.g.~fluctuations in positions and in the speed. Surfers provide an example
of a macroscopic non-thermal suspension where fluctuations are active in origin, and when activity is absent (e.g.~the camphor is exhausted), then the noise is also absent. Crawlers are similar where the activity/noise comes from the vibrating motor. From our measurements, the speed fluctuations have Gaussian behaviour when the particle is far from
the boundaries of the system (\fig{fig:speed-distribution}).
Collisions and interactions
with the boundary  result in a distribution
of speeds near the boundary which has large tails in the case of surfers, but which is qualitatively similar to the bulk distribution for crawlers. Presumably for crawlers this is due to the fact that the collisions with the container are mediated by the surrounding isotropic cup, which is present also when particles are
far from the boundaries and could be responsible for a similar randomization of the speed both in the bulk and close to boundaries as shown in \fig{fig:speed-distribution}.

\subsection{Analytical model}
\begin{figure}[t!]
\centering
\includegraphics[width=0.45\textwidth]{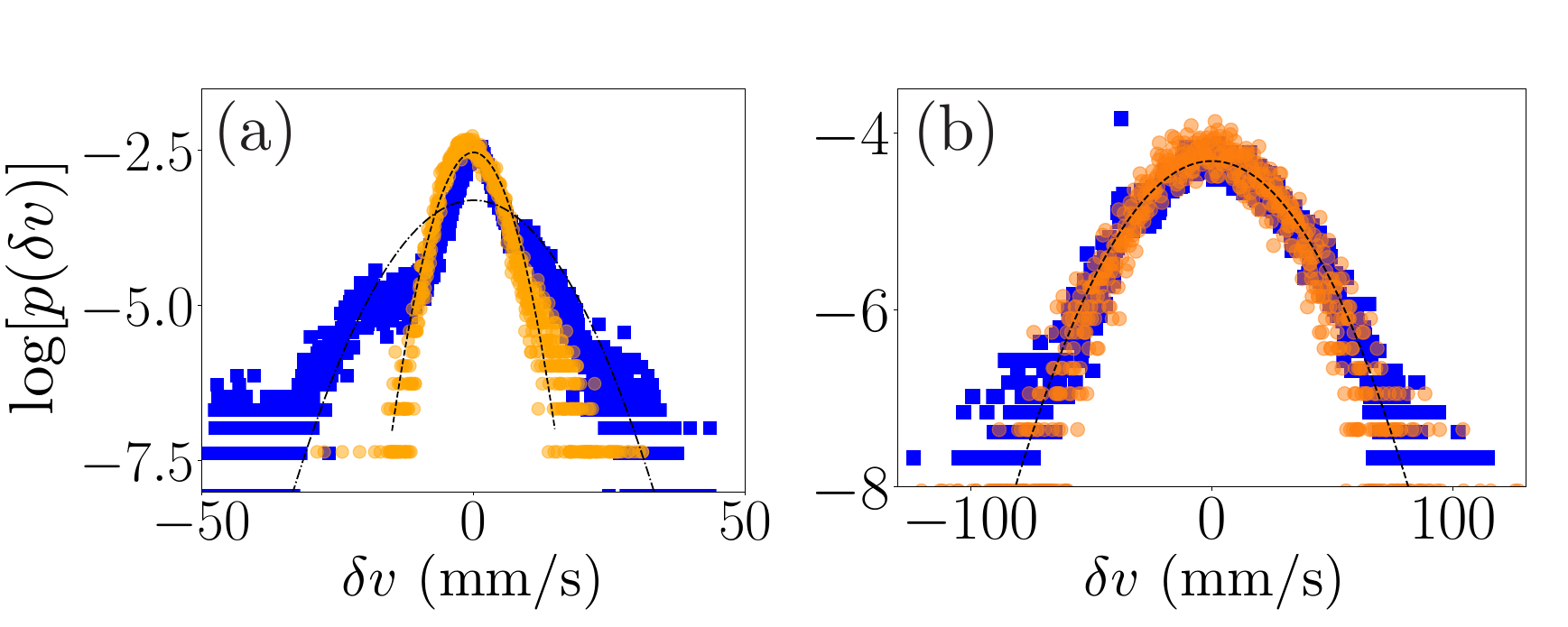}
\caption{Distribution of speed fluctuations, $ \delta v$ (where $\delta v = v - \langle v \rangle$  and $v = \vert \m v \vert$), for surfers (a) and crawlers (b).
Blue squares and orange circles indicate near boundaries and in the bulk, respectively. Fluctuations in the bulk (orange) look nearly Gaussian, while the near the boundary (blue) non-Gaussian tails are evident, particularly for surfers (a). Dashed and dotted-dashed lines show Gaussian distributions with experimental variance.}
\label{fig:speed-distribution}
\end{figure}

To gain insight we examine a simple model similar to  active Brownian particles (ABP) with inertia.
We consider a particle centered at position $\m x$. Then $\dot{\m x } = \m v$
is the particle's velocity whose dynamics is 
 described by
\begin{align}
    & m \dot{\m v} = -\gamma (\m v -\m v_0) -\hh r \frac{\partial U}{\partial r }. \label{eq:vdot}
\end{align}
Here 
$\m v_0$ is the active velocity which  includes both deterministic and fluctuating terms. This is the main difference from the usual ABP model~\cite{romanczuk2012active}: 
in our system
there are no thermal fluctuations  and the noise is directly linked to the activity. 
This is sometimes referred to as active Langevin motion~\cite{lowen2020inertial}.
$  \hh r \frac{\partial U}{\partial r }$ 
is the force ($U$ is the potential) acting on the radial direction, $\hh r$, due to confinement. Writing the 
particle position $ \m x$ in polar coordinates as
$\m x = r (\cos \theta, \sin \theta ) $, then $  \hh r = \m x/r$. 
To compute the radial MSD we 
use the simplest model    to describe the boundary effect: a harmonic potential
$U = k r^2/2$ (see numerical simulations for \textcolor{black}{strong} confinement).
The velocity
vector
$\dot{ \m x}$ is written
$ \dot{\m x} = \dot{r} \hh r + r \dot{\theta} \hh \theta$
and the acceleration
  $ \dot{\m v} = (\ddot{r} - r \dot{\theta}^2 ) \hh r  + (2 \dot{r} \dot{\theta} +  r \ddot{\theta})  \hh \theta$.  
The radial and angular equations are obtained by projecting 
\eq{eq:vdot} onto $\hh r$ and $\hh \theta = (-\sin \theta, \cos \theta) $ respectively.
We note  that these  equations are coupled if inertia is non-negligible, $m \neq 0$.
Projecting the equation on the radial direction, 
we write the active speed
$(\m v_0 \cdot \hh r) = v_0 (1 + \xi(t)) $ as the sum of two terms:
a  constant part, $v_0$, and 
a fluctuating part, $\xi(t) $, which has zero mean and correlations $ \langle \xi(t) \xi(t') \rangle = \Lambda \delta(t-t') $.
 $\Lambda$ describes the strength of
 fluctuations  in the radial velocity, which has a non-thermal origin that can be related to fluctuations in the activity due to chemical reactions for the surfers or the vibrating motor for the crawlers, see \fig{fig:speed-distribution}.

\begin{figure}[t!]
\centering
\includegraphics[width=0.46\textwidth]{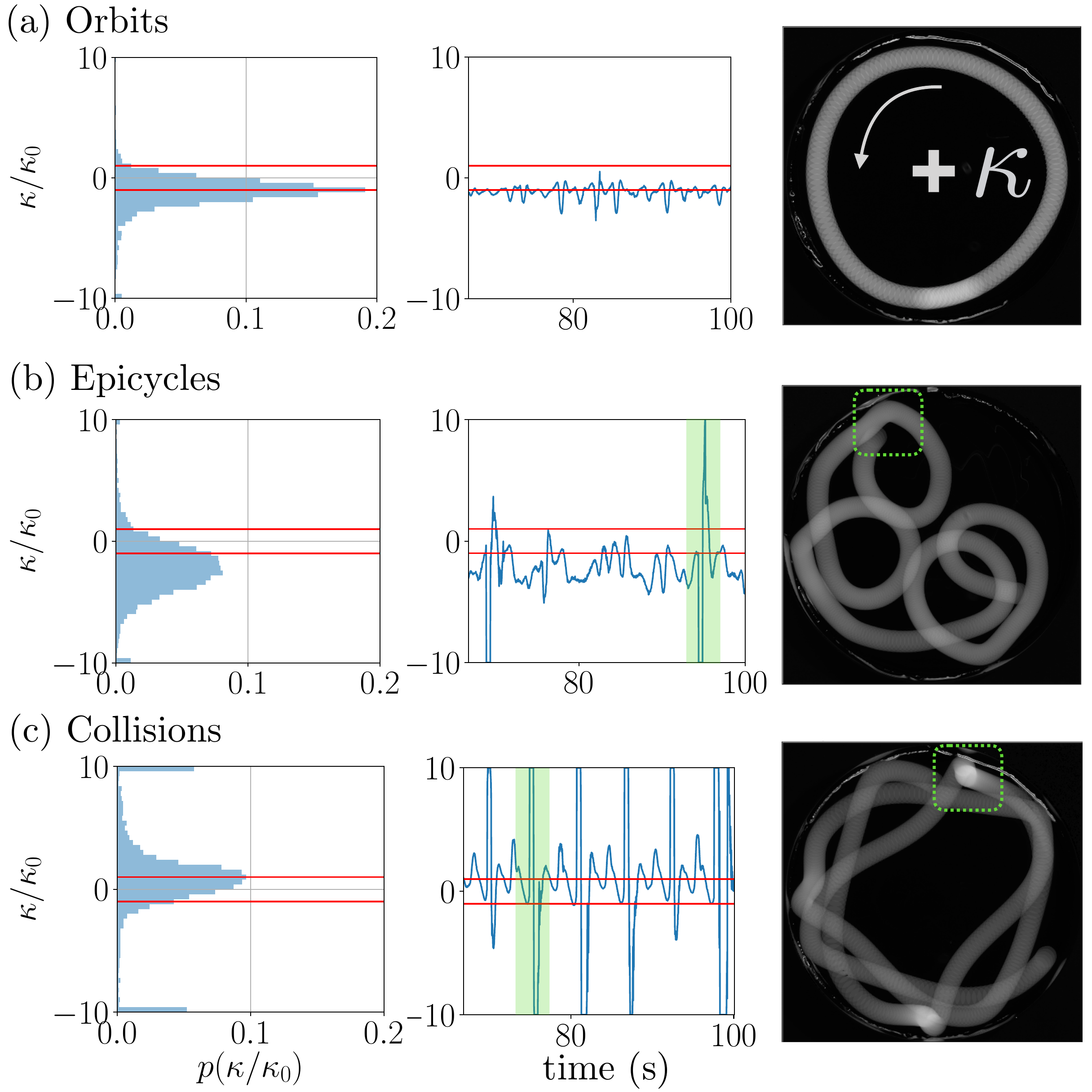}
\caption{Characterization of different dynamical states: (a) orbits, (b) epicycles, and (c) collisions. The left-column shows the distribution of the curvatures. Curvatures are truncated at the value $10\kappa_0$, where $\kappa_0 = 1/R$ is the container curvature. The horizontal red lines indicate the boundary curvature, $\kappa = \pm \kappa_0 $. The middle-column shows the curvature as a function of time, corresponding to the time-lapse trajectories shown in the right-column. A $\kappa / \kappa_0$ value of zero indicates motion in a straight line. (a) Orbiting dynamics exhibit a sharp peak at the curvature corresponding to the boundary, $\vert \kappa_0 \vert$.  (b) Epicycles exhibit a wider distribution of curvatures peaked at a higher value, $\approx \vert 2 \kappa_0 \vert$, indicating ``sharper'' turns.  (c) Collisions exhibit more complex dynamics, with a central peak in curvature similar to the container's boundary, and two large shoulders at very high curvature, $\pm 10 \kappa_0$, associated with collisions where abrupt changes in direction take place.  Shaded green regions in the time dynamics (middle-column) correspond to abrupt changes in curvature indicated in trajectories (right-column).
}
\label{fig:fig4-transitions}
\end{figure}

To study the radial MSD, for simplicity we assume constant angular dynamics, $\dot{ \theta}  = \Omega $. This represents an overall rotational motion, as often observed in experiments (see~\fig{fig:fig4-transitions}), and allows us to decouple the radial equation from the angular one, obtaining
$   
    m \ddot{r}  = - \gamma   \dot{r}  
    + \gamma  v_0 
    + \gamma  v_0 \xi(t) 
    -  m \omega^2_0 r $$
$ 
where $\omega^2_0 = k/m - \Omega^2 $. This result is the classical equation of the
Brownian  oscillator \cite{Uhlenbeck}. 
The 
radial
MSD is computed  (once we subtract the average contribution)from the relation
$
\langle \Delta r^2(t)
 \rangle
=
 \langle [ r(t) -r(0) ]^2 \rangle
= 2 \langle  r^2 \rangle 
-2 \langle  r(t) r(0)  \rangle .
$ 
We obtain,
\begin{align}
  &   
 \langle \Delta r^2(t)
 \rangle
    = 2 \gamma^2 v^2_0 \frac{\Lambda}{m \omega^2_0} 
    \times 
    \nonumber \\
    &
    ( 1
    - e^{-2 \frac{\gamma}{m} t }
    [\cos( \omega_1 t ) + \frac{\gamma}{2 m \omega_1} \sin( \omega_1 t )  ]
    )
      \label{eq:MSDrad}
\end{align}
where $ \omega^2_1 = \omega^2_0 - \frac{\gamma^2}{4 m^2 }  $

The fit reproduces the experimental data accurately for surfers,
see the inset of Fig.\ref{fig:MSDs}. The oscillations
near the plateau could be related to the bouncing dynamics of the particles near the boundary --- which only occurs in the presence of inertia.
In the case of surfers, the fit yields the values
$m/\gamma \sim 40$ s, $\omega_1 = 189.48$ rad/s, and a plateau of $\approx 300$ mm$^2$. For crawlers, we obtain $m/\gamma \sim 5 $ s, and a plateau of $\approx 15000$ mm$^2$, 
with  no appreciable oscillations ($\omega_1 \approx 0.6$ rad/s) see violet curve of \fig{fig:MSDs}(left). These extracted values of $m/\gamma$ suggests that the effect of inertia is stronger in surfers than crawlers in our experimental setup. \textcolor{black}{This is further supported by comparison of $m/\gamma$, the timescale of inertial relaxation, to $\tau \sim R/v_{exp}$, the typical time for a particle to traverse the container. The inertial relaxation time for surfers ($m/\gamma \sim 40$ s) is much larger than timescale to traverse the container ($\tau \sim 0.5$ s); and for crawlers the two timescales are comparable ($m/\gamma \sim 5$ s, and $\tau \sim 5.5$ s).  In both cases the extracted value of $m/\gamma$ indicates inertia is important for the dynamics.  }


How a simple model approximating the confining boundary as a harmonic potential fits the 
MSD
data for both
surfers and crawlers
is not entirely clear. For surfers, presumably interactions with the boundary of the container are mediated by capillary forces
(a meniscus is visible in the proximity of the boundary)~\cite{nakata2015physicochemical}.
To first approximation, a spring-like potential could represent the meniscus force felt by the particle when colliding. A similar effect can be envisaged for the crawlers for interactions of the isotropic cup and container wall.  A difference worth noting is that because the particles are actively driven, the oscillations (albeit damped) do not vanish in the radial MSD.


The equation for the angular coordinate is
$ 
m r \ddot{\theta} = 
-(\gamma r + 2 m \dot{r}) \dot{\theta}  
+
\nu(t)  
$ 
where $\nu(t)$ represents noise, which in the simplest 
case is  
zero except at the instant of collision with the boundary.
As we see, the equation is still coupled to the radial position. To decouple these variables, 
we write $r(t)  \approx r_0$, which represents some average radial position (e.g.~the
peak of the distribution in \fig{fig:radial-distribution}).
Hence, the
equation for the angle is 
$ 
 \ddot{\theta} = 
-\frac{\gamma}{m}    \dot{\theta} 
+
\frac{\nu(t)}{ m r_0 }.  
$ 
The solution for the angle in this case is obtained by 
 considering delta-correlated
noise $\langle  \nu(t') \nu(t'')  \rangle = \Gamma \delta(t'-t'')$, as 
$ \langle \Delta \theta^2 \rangle \approx 
\Omega^2 
t^2$ 
in the regime where inertia dominates~\cite{peliti}. If we neglected $m$,
following
the same calculation  the
angular
MSD would
become
diffusive,
$\langle  \Delta \theta^2 \rangle \approx
\frac{\Gamma }{ \gamma^2 r^2_0  } t $
with $\Gamma$ the strength of fluctuations in angular velocity.  
Together, this analysis suggests the surfers exhibit strong inertial effects because of  the ballistic angular dynamics and oscillations in radial dynamics, while the inertial effects of
our
crawlers~\footnote{
This is not a fundamental difference between crawling vs. surfing.  Other choices of model systems considering different crawlers could yield larger $m/ \gamma $. 
} is less apparent --- consistent with the $m / \gamma$ extracted from the radial analysis. 

\subsection{Dynamical transitions between states}
So far we have discussed statistical
properties of these single particle systems. 
However, this does not distinguish finer
dynamical features that are associated with
a particle's individual trajectory. By focusing on such dynamical details, we find three states:
(1) \emph{Orbits}, where particles move, approximately, on circular trajectories with curvature similar to the size of the confining boundary.
(2) \emph{Epicycles}, where rotation at small scales is coupled to rotation along the container's boundary; 
(3) \emph{Collisions}, where relatively straight trajectories are followed by abrupt changes in direction due to the collisions with the boundary.  While orbits have been observed in soft confinement~\cite{dauchot2019dynamics}, the richer dynamics of epicycles and collisions are unique. This observation, along with our numerical model suggesting transitions between dynamical states is tuned by activity, is the second main result of our \emph{letter}.

 These three dynamical states are shown for surfers in~\fig{fig:fig4-transitions}, and transitions
between them are possible even for a single particle in time.  Signatures of these dynamical states also emerge in the MSDs shown in~\fig{fig:MSDs}.  While all three states have ballistic dynamics at short times, they exhibit different plateau values in the radial dynamics at longer times -- \emph{epicycles} exploring the largest radial area, followed by \emph{collisions} which typically avoid the center of the container, and finally \emph{orbits} where particles typically stay in a small region near the boundary~\fig{fig:MSDs}(left).  These dynamics are also visible in the angular MSDs where \emph{orbits} and \emph{collisions} have the fastest angular motion, and \emph{epicycles} exhibit multi-scale dynamics with comparable motion at short-times but a transition to a slower overall angular motion at long-time due to the local rotations~\fig{fig:MSDs}(right). We show representative examples of timelapse trajectories in~\fig{fig:fig4-transitions}(right).  In crawlers, which have smaller inertial effects ($m/\gamma$),we see similar dynamical states and transitions between, however, less pronounced.  The observation that these states are less pronounced in crawlers suggests that the difference in inertia between surfers and crawlers may play a key role.  We explore this further in the following section using numerical simulations. 

The local curvature $\kappa(t) $ along the particle trajectory
provides a convenient way to characterize the dynamical features. The curvature is computed from the $(x,y)$ coordinates of the particles using a standard 2D formula:  $\kappa(t) = \frac{ (\dot{x}\ddot{y}- \dot{y}\ddot{x}) }{ (\dot{x}^2 + \dot{y}^2)^{\frac{3}{2}}}$.  In interpreting these dynamics, $\kappa(t) > 0$ corresponds to a counter-clockwise motion
and $\kappa(t) < 0$ clockwise. Asymmetry in the distribution of $\kappa / \kappa_0$, for a single trajectory arises due to persistent rotational motion in one direction ($\kappa_0$ is the container curvature). In \fig{fig:fig4-transitions} we show curvature dynamics for our three dynamical states. The \emph{orbits} state (\fig{fig:fig4-transitions}a) is the most straightforward, exhibiting a single relatively narrow peak at a value of $\vert \kappa / \kappa_0 \vert \approx 1$ because the surfer is consistently undergoing rotational motion along the boundary.  The \emph{epicycles} state (\fig{fig:fig4-transitions}b) is mainly characterized by a single large and wide peak.  The peak is wide and biased towards larger values of $\vert \kappa / \kappa_0 \vert$ due to multi-scale rotational dynamics, and thus a wider distribution of curvatures. The \emph{collisions} state (\fig{fig:fig4-transitions}c) is characterized by two large peaks at $\pm 10 \kappa_0$ that correspond to collisions with the boundary generating high-curvature turns and a third more central peak at $\approx \vert \kappa_0 \vert$ due to overall rotating dynamics near the boundary reminiscent of orbits.  It is worth noting there is always an overall rotational motion (i.e. non-zero peak in $\kappa/ \kappa_0$ and ballistic angular MSD).  This means an active particle with non-negligible inertia in a rotationally symmetric container generates persistent rotational motion in one direction.  This is an example of spontaneous breaking of rotational symmetry at the single particle level.

These three dynamical states are observed not only in different realizations of the same experiment, but transitions between states are also observed as a function of time for a single particle.  Since the influence of inertia on a particle does not change in time, this observation suggests that the dynamical states are tuned by some other parameter.  We propose activity tunes the observed dynamical states as we explore in the following section using simulations.



\begin{figure}[t!]
\centering
\includegraphics[width=.5\textwidth]{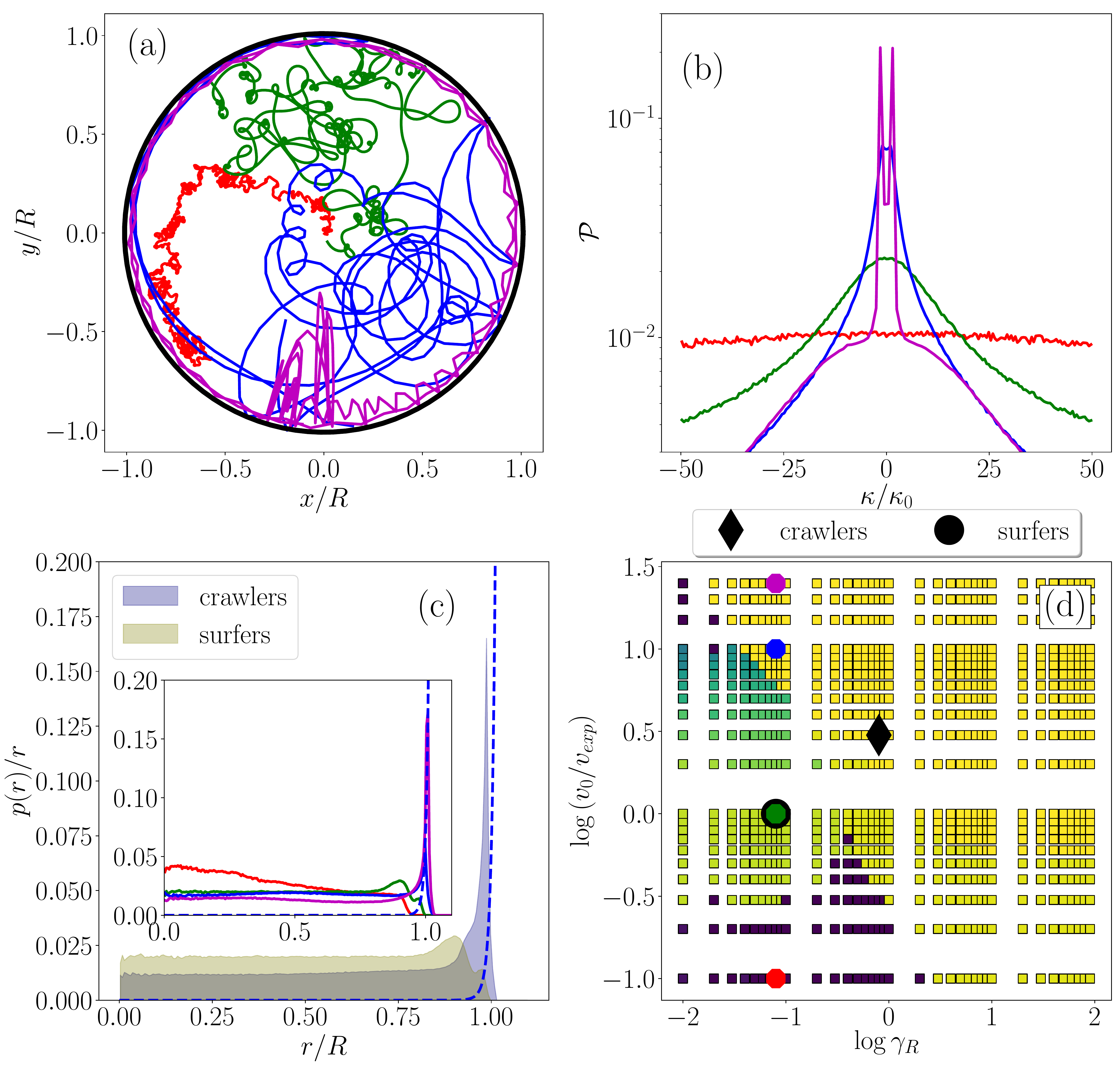}
\caption{ Numerical simulations. 
(a) Representative trajectories of the numerical
model for $v_0 / v_{exp}=0.1,1,10,25$ (red, green, blue, and magenta respectively). (b) Distribution of local curvatures $\mathcal{P}(\kappa / \kappa_0)$, averaged over many trajectories.
(c) The radial distribution function with parameters suitable for reproducing surfers and crawlers.  Inset corresponds to velocities in (a,b). (d) A phase diagram of the position of the peak in the radial distribution function (yellow corresponds to $r/R=1$, violet is $r/R=0$).
Crawlers (black diamond) exhibit a single peaked distribution within but near the boundary. Surfers exhibit a peak well within the boundary with a depletion region near the wall. Red, green, blue, and magenta symbols refer to the location in the phase diagram of the trajectories shown in (a).
} 
\label{fig:sim}
\end{figure}

\subsection{Numerical simulations}
Our simulations show that the dynamical states observed in experiments can be rationalized in terms of a simple model of an active Langevin disc in  the underdamped regime~\cite{lowen2020inertial, scholz2018inertial}. This goes beyond the analytical model described above, by including the dynamic inertial equation for the disc's orientation, $\m n$, and introducing a ``\textcolor{black}{strong}'' confining boundary. In the numerical model, a disc of radius $a$ and mass $m$ moves confined in a circular container of radius $R$. 
 The disc experiences a self-propulsive force along the direction $\m n$ that causes a self-propulsive velocity of magnitude $v_0$. 
 Because of its finite size, the disc is also characterised by it's moment of inertia $I$. The orientation $\m n$ is subjected to a random and uncorrelated noise whose fluctuations are characterized by a rotational diffusion constant \textcolor{black}{$D_r=1$}.

\fig{fig:sim}a shows four representative trajectories at different self-propulsion velocity $v_0 / v_{exp}=0.1,1,10,25$ and fixed $\gamma_R / \gamma=8 \times 10^{-2}$ (the ratio has been chosen to describe the surfers). $v_{exp} \sim 90 $ mm/s is the typical scale of velocity measured in experiments for both surfers and crawlers. As the self-propulsion velocity $v_0$ increases, trajectories transition from Brownian-like to richer dynamical features of \emph{epicycles}, \emph{collisions}, and \emph{orbits}. 
This is an indication that the presence of different regimes observed in experiments may be due to the magnitude of the self-propulsion velocity, $v_0$, i.e., a quantity that that is not directly accessible in experimental measurements because it characterizes only the active process, $\gamma v_0 \mathbf{n}$, and is different from the observed speed, $v_{exp} = \langle\vert\dot{\mathbf{r}}\vert\rangle$.
We look at the statistical distribution of local curvatures $\mathcal{P}(\kappa)$ as shown in \fig{fig:sim}b. At small velocities, trajectories are almost Brownian and the distribution is flat (red). As $v_0$ increases, ballistic dynamics give rise to a peak at $\kappa=0$ (green). When epicycles dominate the dynamics, $\mathcal{P}$ develops a shallow double-peaked structure (blue). The double peak is due to averaging over a large number of trajectories, where in each simulation the disc may be persistently rotating in the positive or negative $\kappa$ direction.  This persistent rotation emerges where collisions with the boundary are observed (blue) at intermediate activity but at higher activity orbits dominate and exhibit a deep double-peaked structure (magenta) as shown in \fig{fig:sim}a,b.



To compare the numerical model with the experiments we look at the steady-state properties through the radial distribution function (using simulation parameters compatible with experiment). The result, shown in \fig{fig:sim}c, is in fair agreement with the experiment. Namely that both surfers and crawlers show an accumulation at a finite distance, $\Delta$, \emph{within} the container boundary, and that this effect is stronger with increasing influence of inertia (e.g.~surfers vs crawlers). Finally, using as a criterion the distance from the boundaries, we develop a phase diagram (\fig{fig:sim}d). The color map indicates the position of the peak of the radial distribution function $r_{peak}$. Yellow indicates that $r_{peak}/ R =1$, violet indicates $r_{peak}/R=0$. The violet region suggests the distribution becomes flat. As one can see, as the inertial effects become negligible (large $\gamma_R$ values), the peak is located at the boundary (yellow) regardless of the self-propulsion speed, as
already
observed in other experiments~\cite{vladescu2014filling, takatori2016acoustic}. However, when inertia dominates (at lower values of $\gamma_R$), the phase diagram shows rich behavior as the self-propulsion speed is varied.  Very low $v_0$ shows uniform distributions (violet) as expected for Brownian motion, and higher $v_0$ show a peak of the distribution at a finite distance within the boundary (shades of green), $\Delta$, as observed in our experiments --- e.g.~\emph{epicycles} and \emph{collisions} tend to move the peak away from the boundary.  \textcolor{black}{Altogether, our simulations suggest that an inertial active disc under strong confinement can dynamically transition from Brownian-like to \emph{epicycles}, \emph{collisions}, and \emph{orbits} by increasing the self-propulsion velocity, $v_0$. }

It is worth noting that the structure of the radial distribution function shows richer features than in the overdamped case~\cite{vladescu2014filling}.  This is made evident by the green regions in the phase diagram (upper-left) where the peak of the distribution not only moves away from the boundary, but it also exhibits a depletion region close to the wall --- visible in \fig{fig:sim}c. This observation is in agreement with experiments, most visibly in surfers (green curve in \fig{fig:radial-distribution}) that exhibit a peak
at finite distance $\Delta $ from the boundary. Moreover, the distributions in both experiments and simulations exhibit a ``shoulder'' in the depletion region (green curves in \fig{fig:radial-distribution} and \fig{fig:sim}c).

\section{Conclusions}
To summarize --- In ``\textcolor{black}{strong}'' confinement a Brownian particle will uniformly explore the space; and an overdamped active particle will accumulate at the container wall. In this article,
we show that a self-propelled particle with non-negligible inertia gives rise to two new effects: (1) Particles accumulate at a finite distance \emph{within} the container wall, and this distance increases with activity and inertia. (2) Three dynamical states (and transitions between them) are observed that can be characterized by the local curvature, all of which include breaking of rotational symmetry.  Both (1) accumulation and (2) transitions between  dynamical states can be tuned by activity, $v_0$, and only exist when inertia is non-negligible.  These observations open a new 
avenue  for inertial active matter, because they show that, thanks to inertial effects, active particles can be spatially sorted in target regions by varying the properties of the particle itself without introducing an external field or opportune sculpured environments \cite{galajda2007wall,koumakis2013targeted}.
%
Further modeling and experiments are necessary to fully understand the role of inertia on the spatial distribution of active particles, the finer features of their dynamical states, and how this affects multi-particle interactions.


{\it  Acknowledgments.---} MP acknowledges funding from Regione Lazio, Grant Prot.~n.~85-2017-15257 (”Progetti di Gruppi di Ricerca - Legge 13/2008 - art.~4”). MP is supported by the H2020 program under the MSCA grant agreement No.~801370 and by the Secretary of Universities and Research of the
Government of Catalonia through Beatriu de Pin\'os program Grant No.~BP 00088 (2018). SE and AE acknowledge the Black Family Fellowship and LSAMP NSF grant HRD-1826490.  WWA acknowledges funding from CSUF RSCA and NSF.  This material is based upon work supported by the National Science Foundation under grant no.~DMS-2010018 and CHE-1560390.

\bibliographystyle{unsrt}
\bibliography{biblio.bib}


\end{document}